\documentclass{article}
\usepackage{amsmath}
\usepackage{bbm}
\usepackage{amsfonts}
\usepackage{dsfont}
\usepackage{amssymb}
\usepackage{mathrsfs}
\usepackage{calrsfs}
\usepackage{color}
\usepackage{epsfig}
\usepackage{graphicx}

\usepackage[super,square,comma,sort&compress]{natbib}
\usepackage[pagebackref,colorlinks,linkcolor=blue, citecolor=blue, hyperindex,pdfstartview=FitH,plainpages=false,CJKbookmarks]{hyperref}

\def\bc{\begin{center}}

\def\ec{\end{center}}
\def\be{\begin{eqnarray}}
\def\ee{\end{eqnarray}}
\def\bde{\begin{definition}}
\def\ede{\end{definition}}
\def\bpro{\begin{proof}}
\def\epro{\end{proof}}
\def\bth{\begin{theorem}}
\def\eth{\end{theorem}}
\def\ble{\begin{lemma}}
\def\ele{\end{lemma}}
\def\bpr{\begin{proposition}}
\def\epr{\end{proposition}}
\def\bco{\begin{corollary}}
\def\eco{\end{corollary}}
\numberwithin{equation}{subsection}

\newcommand{\omits}[1]{}

\definecolor{dyellow}{rgb}{1.,0.8,.0}
\definecolor{myblue}{rgb}{.1,.1,.7}
\definecolor{dcyan}{rgb}{.0,.6,.6}
\definecolor{dmagenta}{rgb}{0.6,0.0,0.6}
\definecolor{brown}{rgb}{0.6,0.2,0.}
\definecolor{darkblue}{rgb}{.0,.0,0.5}
\definecolor{darkred}{rgb}{0.75,0.0,0.0}
\definecolor{orange}{rgb}{1.,.6,.0}
\definecolor{dorange}{rgb}{0.8,.4,.0}
\definecolor{darkgreen}{rgb}{0.0,0.6,0.0}
\definecolor{purple}{rgb}{.4,.0,.4}

\renewcommand{\abstract}[1]{\par\noindent{\small{\bf Abstract\/}: #1}}
\newenvironment{proof}[1][Proof.]{\begin{trivlist}
\item[\hskip \labelsep {\bfseries #1}]}{\end{trivlist}}

\newenvironment{remark}[1][Remark]{\begin{trivlist}
\item[\hskip \labelsep {\bfseries #1}]}{\end{trivlist}}

\newcommand{\qed}{\nobreak \ifvmode \relax \else
     \ifdim\lastskip<1.5em \hskip-\lastskip
      \hskip1.5em plus0em minus0.5em \fi \nobreak
      \vrule height0.75em width0.5em depth0.25em\fi}
\newtheorem{theorem}{Theorem}[subsection]
\newtheorem{lemma}[theorem]{Lemma}
\newtheorem{definition}[theorem]{Definition}
\newtheorem{proposition}[theorem]{Proposition}
\newtheorem{corollary}[theorem]{Corollary}

\newcommand\btd{\raise 2pt
\hbox{$\hat\bigtriangledown$}\hskip 1.5pt}
\newcommand\bt{\raise 2pt
\hbox{$\bigtriangledown$}\hskip 1.5pt}

\numberwithin{equation}{subsection}
\begin{document}
\title{Noncommutative differential calculi and the unifying zero curvature representation of integrable systems
}
\date{  }
\author{
\small{Yongqiang Bai $^{a, b }$,
Ming Pei $^{b}$
, Huijuan Fu$^{b}$}\\
\small{$^{a}$ Institute of Contemporary Mathematics, Henan
University, Kaifeng
475004, China }\\
\small{$^{b}$School of Mathematics and Information Science,
Henan University, Kaifeng 475004, China}\\
}
\maketitle
\begin{abstract}
{Derivation-based differential calculi are of great importance in noncommutative geometry, noncommutative gauge theory and integrable systems. In this paper, we propose the connection and curvature from a class of deformed derivation-based differential calculus. By means of this theory,  we give out the zero-curvature representation of the continuum-continuum, discrete-continuum and discrete-discrete integrable systems in an unifying manner.}
\end{abstract}
\begin{flushleft}
\end{flushleft}
\tableofcontents
\section{Introduction}
\label{intro}
Noncommutative differential geometry (NCDG) \cite{Connes1, Connes2} and other noncommutative differential calculi (NCDC)
have arisen much interests for their  theoretical developments and applications in string theory, gauge theory, quantum field theory, the models with Higgs field , the matrix theory \cite{Connes2, Cha, Madore, Seiberg} and so forth.

Theories of connection and curvature are also essential ingredients in classical differential geometry. They play important roles in gauge fields theory, general relativity, integrable systems. From the mathematical point of view, the fundamental forces in nature are curvatures of some appropriate fiber bundles.  Recently, these theories admit some noncommutative versions\cite{Connes1, Connes2, Madore, Dubois1, Dubois2, Dubois3, Dubois4, Dubois5, Dubois6, Dubois7, Eric, Wu1, Dubois8}, which are all in close related to the differential calculi one considers. These noncommutative differential calculi are all based on the J. L. Koszul algebraic approach to differential geometry\cite{Koszul}.
In general, the generation of classical differential calculi to noncommutative case are not unique and actually the extension of commutative tools to noncommutative setting is never straightforward. In the procedure, one must have many examples and their applications in mind. Otherwise, if it is just a pro forma generation, the theory will be completely meaningless.

In\cite{Koszul}, J.L. Koszul described a powerful algebraic version of differential geometry in terms of a commutative associative algebra $\mathcal{C}$, $\mathcal{C}$-modules and connections on these modules. For the application to differential geometry, $\mathcal{C}$ is the algebra of smooth functions on a manifold and elements of the $\mathcal{C}$-module are sections of smooth vector bundles over the manifold. The fact that classical differential geometry admits such an algebraic formulation is at the very origin of the idea of noncommutative differential geometry.

In noncommutative geometry, one replaces the commutative associative algebra $\mathcal{C}$ by an associative algebra
$\mathcal{A}$ which is not necessary commutative.  The ordinary exterior differential calculi $\Omega$ is replaced by a differential calculus over $\mathcal{A}$ such that $ \Omega^{0}=\mathcal{A}$ with a differential operator denoted by $d$.

In \cite{Wu1}, the authors have proposed a discrete fiber bundle, connection and curvature theory in a way similar  to the continuous case. We will show that their theory is actually a concrete realization of the noncommutative differential calculi basing on a deformed derivative. We also propose the general differential-difference theory and consider their applications in continuous and discrete integrable systems.

The organization of this paper is as follows. In section 2,  the noncommutative geometry theory is reviewed briefly.
In section 3 and 4, we present a concrete realization of the noncommutative connection and curvature theory. We show that theory in \cite{Wu1} is a special case of this theory.
In section 5, the application of the theory to integrable systems is considered. We give out three important integrable equations as our examples. Finally, in section 6, we end this paper with some conclusions.
\section{ Noncommutative geometry from global algebraic point of view}
In this section, we briefly recall the noncommutative differential calculi theory and summarize the properties of derivative-based differential calculi needed throughout this paper. This will fix our notations and conventions. We refer to \cite{Dubois8, Dubois9} for more details.

\subsection{Differential calculus on an associative algebra  $  \mathcal{A} $}

Let  $  \mathcal{A} $ be an associative complex algebra(not necessary commutative) with unit $\mathbbm{1}$  and $\mathcal{Z}$$ (\mathcal{A})$ be the center of $\mathcal{A}$.
\bde
The vector space of derivatives of $\mathcal{A}$ is the space
\be
Der(\mathcal{A})=\{\mathfrak{X}:\mathcal{A}\rightarrow \mathcal{A}\mid \mathfrak{X} ~is~linear, \mathfrak{X}(ab)=\mathfrak{X}(a)b+a\mathfrak{X}(b)\}.
\ee
\ede
$Der(\mathcal{A})$ is a Lie algebra for the bracket
$[\mathfrak{X},\mathfrak{Y} ]a=\mathfrak{X}\mathfrak{Y}a-\mathfrak{Y}\mathfrak{X}a, $ and a $\mathcal{Z}$$ (\mathcal{A})$-module for the product $(f\mathfrak{X})a=f(\mathfrak{X}a). $
By means of this definition and properties of the space $Der(\mathcal{A})$, one can obtain the graded differential calculus on the algebra  $\mathcal{A}$.

Let $\underline{\Omega}^{n}_{Der}(\mathcal{A})$ be the set of $\mathcal{Z} (\mathcal{A})-$multilinear antisymmetric maps from $Der(\mathcal{A})^{n}$ to $\mathcal{A},$ with $\underline{\Omega}^{0}_{Der}(\mathcal{A})=\mathcal{A},$ and let
\be
\underline{\Omega}^{\bullet}_{Der}(\mathcal{A})=\oplus_{n\geq 0}\underline{\Omega}^{n}_{Der}(\mathcal{A}).
\ee
The space $\underline{\Omega}^{\bullet}_{Der}(\mathcal{A})$ can be equipped with a structure of $\mathds{N}$-graded  algebra by means of the product
\be
(\omega\wedge\eta)(\mathfrak{X}_{1},\ldots, \mathfrak{X}_{p+q})&=&\frac{1}{p!q!}\sum_{\sigma\in \mathfrak{G}_{p+q}}(-1)^{sign(\sigma)}\omega(\mathfrak{X}_{\sigma(1)},\ldots, \mathfrak{X}_{\sigma(p)})\nonumber\\
&&\eta(\mathfrak{X}_{\sigma(p+1)},\ldots,\mathfrak{X}_{\sigma(p+q)}).
\ee
\bde
Let $(\underline{\Omega}^{\bullet}_{Der}(\mathcal{A}), \wedge)$ be a graded algebra with $\underline{\Omega}^{0}_{Der}(\mathcal{A})=\mathcal{A}$, one can get the differential calculus on the algebra $\mathcal{A} $ by defining the differential operator $d$ of degree 1 using the Koszul formula
\begin{eqnarray}
d\omega(\mathfrak{X}_{1}, \ldots, \mathfrak{X}_{n+1})&=&\sum_{i=1}^{n+1}(-1)^{i+1}\mathfrak{X}_{i}\omega(\mathfrak{X}_{1}, \ldots, \hat{\mathfrak{X}}_{i}, \ldots, \mathfrak{X}_{n+1})+\nonumber\\
&&\sum_{i<j}(-1)^{i+j}\omega([\mathfrak{X}_{i},\mathfrak{X}_{j}],  \ldots, \hat{\mathfrak{X}}_{i}, \ldots, \hat{\mathfrak{X}}_{j}, \ldots, \mathfrak{X}_{n+1}).
\end{eqnarray}
where $\hat{\mathfrak{X}}_{i}$ means as usual omission of the $i$th argument $\mathfrak{X}_{i}$.
\ede
There are many ways to define a differential calculus over an associated algebra. In concrete applications, the differential calculus should be construct to adapt to the structures and to the tools in the noncommutative space\cite{Masson1}.
\subsection{Noncommutative connection and curvature}
The theory of noncommutative connection depend essentially on the derivative-based differential calculus and looks very similar to the ordinary connection.
Every theory of connection relies on modules over an associative algebra. In general, one take the left or right finite projective module or consider the bimodule.
\bde
Let $\mathcal{A}$ be an associated algebra,   $( \underline{\Omega}^{\bullet}_{Der}(\mathcal{A}), d)$ be a differential calculus over $\mathcal{A}$, $\mathcal{M}$ be a left $\mathcal{A}$-module , then a left $\Omega$-connection on $\mathcal{M}$ is a linear mapping
$$
\mathcal{D}: \mathcal{M}\rightarrow \underline{\Omega}^{1}_{Der}(\mathcal{A})\otimes_{\mathcal{A}}\mathcal{M}
$$
such that one has
\be
\mathcal{D}(am)=a\mathcal{D}(m)+d(a)\otimes_{\mathcal{A}}m
\ee
for all $a\in \mathcal{A}, m\in \mathcal{M}.$
\ede
One can extends $\mathcal{D}$ to $\underline{\Omega}^{\bullet}_{Der}(\mathcal{A})\otimes_{\mathcal{A}}\mathcal{M}$ by setting
\be
\mathcal{D}(\omega \otimes_{\mathcal{A}}m)=(-1)^{n}\omega \mathcal{D}(m)+d(\omega)\otimes_{\mathcal{A}}m
\ee
for
$
\omega\in \underline{\Omega}^{n}_{Der}(\mathcal{A})
$
and
$
m\in \mathcal{M}.
$
\bde
The curvature of the connection $\mathcal{D}$ is just the restriction of $\mathcal{D}^{2}$ to $\mathcal{M}$:
\be
\mathcal{D}^{2}:  \mathcal{M}\rightarrow \underline{\Omega}^{2}_{Der}(\mathcal{A})\otimes_{\mathcal{A}}\mathcal{M}.
\ee
\ede
In the noncommutative case, we can also define the bimodule connection\cite{Dubois8, Dubois9}.
\bde
A bimodule connection on an $\mathcal{A}$-bimodule $\mathcal{M}$ is a triple $ (\mathcal{M}, \mathcal{D}, \sigma )$, where $\mathcal{D}: \mathcal{M}\rightarrow \underline{\Omega}^{1}_{Der}(\mathcal{A})\otimes_{\mathcal{A}}\mathcal{M}$ is a left $\mathcal{A}$-connection, and
$\sigma: \mathcal{M}\otimes_{\mathcal{A}}\underline{\Omega}^{1}_{Der}(\mathcal{A})\rightarrow \underline{\Omega}^{1}_{Der}(\mathcal{A})\otimes_{\mathcal{A}}\mathcal{M}$ is a bimodule map with the property:
\be
\mathcal{D}(ma)=\mathcal{D}(m)a+\sigma(m\otimes da),~~~\forall m\in \mathcal{M}, a\in \mathcal{A}.
\ee
\ede
\section{ The concrete realization on the algebra $\mathcal{A}=C^{\infty}(L^{p}\times R^{q})$ }
\label{sec:3}
In this section, we will consider a concrete realization of the noncommutative connection and curvature presented in \S 2 based on a deformed derivative.
We use $L^{p}$ to denote the p-dimensional lattice space with the same equal space $1$ and consider the algebra $\mathcal{A}=C^{\infty}(L^{p}\times R^{q})$.
\subsection { The corresponding N-graded algebra $\Omega^{\bullet}$ }
\label{subsec:3.1}
Let N denote the semi-discrete space $N=L^{p}\times \mathbb{R}^{q},$ where
$L^{p}$ is the p-dimensional lattice space and $\mathbb{R}^{q}$ is
the q-dimensional Euclidean space. For convenience, we denote the semi-discrete space N by
$$
 N=L^{p}\times
\mathbb{R}^{q}=\{(\vec{n}, \vec{x})=(n^{1},\cdots,n^{p},x^{1},\cdots,x^{q})|
n^{\mu}\in \mathbb{Z}, x^{i}\in \mathbb{R}, 1\leq \mu\leq p, 1\leq i\leq
q\}
$$
Let $\mathcal{A}=C^{\infty}(N)=C^{\infty}(L^{p}\times R^{q}).$
\bde
For any function $f(\vec{n}, \vec{x})\in
\mathcal{A}$, we define the differences of $f$ in the $\mu$-th
$(1\leq\mu\leq p)$ direction as
\begin{equation}
\triangle_{\mu}f(\vec{n}, \vec{x})=\triangle_{\mu}f(n^{1},\cdots,n^{\mu},\cdots,n^{p}
,x^{1},\cdots,x^{j} ,\cdots ,x^{q})=(E_{\mu}-id)f,
\end{equation}
where $E_{\mu}$ is the forward shift operator in the $\mu$-
direction
\begin{equation}
E_{\mu}f(\vec{n}, \vec{x})=f(\vec{n}+\hat{\mu}, \vec{x}), ~~~\hat{\mu}=( \underbrace{0, \cdots, \overbrace{1}^{\mu},\cdots, 0}_{p} ).
\end{equation}
\ede
Similar to the continuous case, we can define
the differential form and exterior differential operator in the
semi-discrete case.
\bde
Let $(\vec{n}, \vec{x} )$ be any point in the
semi-discrete space $N=L^{p}\times \mathbb{R}^{q}$, $\Delta_{\mu}$ be
the difference operator defined above,
$
\partial_{i}
$
 be the ordinal partial differential operator. The tangent space  $T_{(\vec{n},\vec{ x})}(N) $ at the point $(\vec{n}, \vec{x} ) $ is:
$$
T_{(\vec{n},\vec{ x})}(N) =
Span\{\Delta_{1},\cdots,\Delta_{p},\partial _{1},\cdots, \partial_{q}\}.
$$
\ede
\begin{remark}.
It should be noted that for any two functions $f(\vec{n}, \vec{x}), g(\vec{n}, \vec{x})\in N=L^{p}\times
\mathbb{R}^{q},$ one have the following relations for the generators of $T_{(\vec{n},\vec{ x})}(N)$:
\be
\partial_{i} (fg)&=&\partial_{i} (f)g+f\partial_{i} (g),\;i=1, \cdots, q,\\
\Delta_{\mu} (fg)&=&\Delta_{\mu} (f)E_{\mu}g+f \Delta_{\mu} (g),\; \mu=1, \cdots, p.
\ee
\end{remark}
From Eq. (3.1.4), one know that the tangent space or the derivative space is a deformed one. The generator $\partial_{i}$ obey the ordinary Leibniz rule, however the generator $\Delta_{\mu}$ obey a deformed Leibniz rule.
\bde
The cotangent space  $T^{\ast}_{(\vec{n}, \vec{x})}(N)$ at the point $(\vec{n}, \vec{x} ) $ is the dual space of $T_{(\vec{n}, \vec{x})}(N) $ :
$$
T^{\ast}_{(\vec{n}, \vec{x})}(N)=
Span\{dn^{1},\cdots,dn^{\mu},\cdots,dn^{p},dx^{1},\cdots,dx^{i},
\cdots,dx^{q} \}.
$$
with the following relations
\be
\left\{
\begin{array}{llllllll}
\langle dn^{\mu},\Delta_{\nu}\rangle=\delta^{\mu}_{\nu},\\
\langle dn^{\mu}, \partial_{j}\rangle=0,\\
\langle\ dx ^{j},\Delta_{\mu}\rangle=0,\\
\langle\ dx^{j}, \partial_{i}\rangle=\delta^{j}_{i},\\
dn^{\mu}\wedge dn^{\mu}=0,\\
dn^{\mu}\wedge dn^{\nu}=-dn^{\nu}\wedge dn^{\mu},\ \ \
\ \ \ \mu \neq \nu,\\
dn_{\mu}\wedge dx ^{i}=-dx^{i}\wedge dn_{\mu},\\
dx^{i}\wedge dx^{i}=0,\\
dx^{i}\wedge dx^{j}=-dx^{j}\wedge dx^{i},\ \ \ \ \ \ i\neq
j,\\
{[dx^{i}, f(\vec{n}, \vec{x})]}=0, i=1,\cdots, q,\\
{[dn^{\mu}, f(\vec{n}, \vec{x})]}= \triangle_{\mu}f(\vec{n}, \vec{x}) dn^{\mu}, \mu=1,\cdots, p.
\end{array}\right.
\ee
\ede
We can therefore define the tangent bundle and cotangent bundle over
$N=L^{p}\times R^{q}$ as:
\begin{equation}
T(N):=\cup_{(\vec{n}, \vec{x})\in N}
T_{(\vec{n},\vec{ x})}(N)
\end{equation}
and
\begin{equation}
T^{\ast}(N):=\cup_{(\vec{n}, \vec{x})\in N}
T^{\ast}_{(\vec{n},\vec{ x})}(N).
\end{equation}

From (3.1.5) we know that the left $\mathcal{A}$-module $\Omega^{\ast}$ and the right
$\mathcal{A}$-module $\Omega^{\ast}$  are not the same. We will only consider the left $\mathcal{A}$-module $\Omega^{\ast}.$  However, for the base space is $N=L^{p}\times \mathbb{R}^{q}$,
the algebra is
$
\mathcal{A}=C^{\infty}(N).
$ The center of the algebra $\mathcal{A}$ is itself: $\mathcal{Z}(\mathcal{A}) =\mathcal{A}.$ If we denote also by ${\Omega}^{\ast}_{Der}(\mathcal{A}) \subset \underline{\Omega}^{n}_{Der}(\mathcal{A})$ the sub differential graded algebra generated in degree 0 by $\mathcal{A}$, i.e., each element in ${\Omega}^{n}_{Der}(\mathcal{A})$ is a sum of the form $ f dn^{1}\wedge \cdots\wedge dn^{p}\wedge dx^{1}\wedge
\cdots\wedge dx^{q}$ for $p+q=n, f\in C^{\infty}(N).$  One have
$
\Omega^{\ast}_{Der}(\mathcal{A}) = \underline{\Omega}^{\ast}_{Der}(\mathcal{A})=\Omega^{\ast}(\mathcal{A} )
$
in this case.

We can thus construct the whole exterior algebra
$\Omega^{\ast}=\oplus_{m\in Z} \Omega^{m} $ on
$T^{\ast}(N)$ with bases
$$\underbrace{1}_{0-form}, \underbrace{dn ^{\mu}, dx^{i}}_{1-forms}, \underbrace{dn^{\mu}\wedge dn^{\nu},
dn^{\mu}\wedge dx^{i}, dx^{i}\wedge dx^{j}}_{2-forms}
,\cdots, \underbrace{dn^{1}\wedge \cdots\wedge dn^{p}\wedge dx^{1}\wedge
\cdots\wedge dx^{q}}_{(p+q)-form}.$$

If $\omega\in \Omega^{r}(\mathcal{A})$ be a $r$-form ($ 0\leq r\leq
p+q$), then $\omega$ may be written as
\begin{eqnarray*}
\omega&=&\sum_{\mu_{1}\mu_{2}\cdots \mu_{k}i_{1}i_{2}\cdots i_{l} \atop
k\leq p,l\leq q, k+l=r} f_{\mu_{1}\mu_{2}\cdots \mu_{k}i_{1}i_{2}\cdots
i_{l}}(\vec{n}, \vec{x})\\
&&
dn^{\mu_{1}}\wedge dn^{\mu_{2}}\wedge\cdots\wedge dn^{\mu_{k}}\wedge
dx^{i_{1}}\wedge dx^{i_{2}}\wedge\cdots\wedge dx^{i_{l}}.
\end{eqnarray*}

For notation simplicity, we will also write $\omega$ as
$$\omega=\sum_{IJ} f_{IJ}dn^{I}\wedge dx^{J}.$$
We will define the semi-discrete exterior derivative operator $d$ in the following so that
$\Omega^{\ast}(\mathcal{A})$ can constitute a graded differential algebra.
 \bde
 For the algebra
$\Omega^{\ast}(\mathcal{A})$, the semi-discrete exterior differential operator $d$ is: \\
\centerline{$d:\Omega^r(\mathcal{A})\to\Omega^{r+1}(\mathcal{A})$}

$1.$ If $f(\vec{n}, \vec{x})\in\Omega^{0}(\mathcal{A})=\mathcal{A}$, then
$df=d_{D}f+d_{C}f,$ where $d_{D}f$ and $d_{C}f$  are defined respectively as
$$d_Df= \sum_{\mu=1}^{p}\Delta _\mu f dn^{\mu}$$
$$d_Cf=\sum_{i=1}^{q}\partial_{i} f dx^i$$

 $2.$ If
$$\omega=\sum_{IJ}f_{IJ}\chi^I \wedge dx^J\  \in \ \Omega^r
(\mathcal{A}),$$  then
\begin{equation}
d\omega=\sum_{IJ} df_{IJ}\wedge\chi^I\wedge dx^J.
\end{equation}
\ede For convenience, we will refer the operator $d_{D}$ as the
discrete differential operator and $d_{C}$ as the continuous
differential operator respectively. It is obvious that the
semi-discrete exterior differential operator $d$ is the sum of the
discrete differential operator $d_{D}$ and continuous differential
operator $d_{C},$ i.e.,
\begin{equation}
d=d_{D}+d_{C}.
\end{equation}
From which we know that
$$d(dn^{\mu})=0,\  d(dx^{j})=0$$
\bpr Let $d$ be a semi-discrete exterior differential operator, it has the following properties:
$$
\begin{array}{ll}
1)& d(\omega\wedge\tau)=(d\omega)\wedge\tau+(-1)^{\deg\omega}\omega\wedge d\tau ,\\
2)& d^{2}=0 ,\\
3)& (df)(v)=v(f), ~~~\forall f\in\Omega^{0}(\mathcal{A}),v\in T(N).
\end{array}
$$
\epr
It is obvious that the algebra $\Omega^{\ast}(\mathcal{A})$ is a graded differential algebra and may be expressed as
$$\Omega^{\ast}(\mathcal{A})=\mathop  \oplus
\limits_{r= 0}^{p+q}\Omega^r(\mathcal{A}).$$ where
$\Omega^{r}(\mathcal{A})$ is consist of smooth
r-forms defined on $L^{p}\times
\mathbb{R}^{q}$ and $\Omega^{0}(\mathcal{A})=\mathcal{A}.$
\subsection {Semi-discrete complex and its exactness}
\label{subsec:3.2}
Similar to the continuous case, one can also obtain a exact complex in the semi-discrete space $N=L^{p}\times
\mathbb{R}^{q}.$
\bpr
The complex
$$0 \rightarrow \mathbb{R} \xrightarrow{i}
\Omega^{0}(\mathcal{A}) \xrightarrow{d}
\Omega^{1}(\mathcal{A}) \xrightarrow{d}
\Omega^{2}(\mathcal{A}) \xrightarrow{d} \cdots
\xrightarrow{d} \Omega^{p+q}(\mathcal{A}) \rightarrow
0$$ is exact.
\epr
\bpro
See details in \cite{Fu, lb1, lb2} \qed
\epro
\section{The noncommutative semi-discrete connection and curvature}
In this section, we will use the similar symbols and notation as in \cite{Wu1} to obtain the semi-discrete bundle, connection and curvature.
 \subsection{semi-discrete bundle and semi-discrete sections}
If $(P, \mathfrak{p}, M, G) $ is a principal $G-$bundle over a (p+q)-dimensional base manifold $M$ with the structure group $G$.
Let the base manifold $M$ be partial discretized as $M\simeq N=L^{p}\times \mathbb{R}^{q}$.
 For any point $(\vec{n}, \vec{x})\in N$, the corresponding fiber $\mathfrak{p}^{-1}(\vec{n}, \vec{x})$ is isomorphic to the structure group $G$. The semi-discrete principal bundle $Q(N , G)$ will be defined as the union of all these fibers:
 \be
 Q(N , G)=\bigcup\nolimits_{(\vec{n}, \vec{x})\in N}\mathfrak{p}^{-1}(\vec{n}, \vec{x}).
 \ee
 If the structure group $G$ is a linear matrix group $GL(m, \mathds{R})$, one can get the associated vector bundle $V=V(N, \mathds{R}^{m}, GL(m, \mathds{R}))$. In this case, for any $(\vec{n}, \vec{x})\in N ,$ the fiber $F_{(\vec{n}, \vec{x})}= \mathfrak{p}^{-1}(\vec{n}, \vec{x})$ is isomorphic to an linear space $\mathds{R}^{m} $ with the right action of $GL(m, \mathds{R})$ on it. The semi-discrete vector bundle $V(N, GL(m, \mathds{R})) $ is defined as:
 \be
 V(N, GL(m, \mathds{R}))=\bigcup\nolimits_{(\vec{n}, \vec{x})\in N}\mathfrak{p}^{-1}(\vec{n}, \vec{x}).
 \ee

 The semi-discrete section on a semi-discrete principle bundle is a map as:
 \be
 s: N=L^{p}\times \mathbb{R}^{q}=\{(\vec{n}, \vec{x} )\}\rightarrow Q(N, G), ~~s(\vec{n}, \vec{x} )\in G,
 \ee
 Similarly, the semi-discrete section on a semi-discrete vector bundle is a map as,
 \be
 s: N=L^{p}\times \mathbb{R}^{q}=\{(\vec{n}, \vec{x} )\}\rightarrow V(N, GL(m, \mathds{R})), ~~s(\vec{n}, \vec{x} )\in \mathds{R}^{m}.
 \ee
 We also use $\Gamma(Q),  \Gamma(V)$ to denote the set of the semi-discrete sections on the corresponding principle bundle and the corresponding vector bundle respectively.

 The tangent bundle $TN$ and the cotangent bundle  $T^{\ast}N$ defined in \S 3.1 are two examples of semi-discrete vector bundles over the space $N=L^{p}\times \mathbb{R}^{q}.$ The cotangent bundle  $T^{\ast}N$ is of great importance in defining the connection and curvature in the following sections.
 \subsection{Noncommutative semi-discrete connection}
 \bde
 A left $\mathcal{A}$-module difference-differential connection or covariant difference-differential derivative is a linear map:
 \be
 \mathcal{D}: \Gamma(V)\rightarrow \Gamma(T^{\ast}N)\otimes \Gamma(V)
 \ee
 which satisfies the following relations:
 \be
 && \mathcal{D}(s_{1}+s_{2})=\mathcal{D}(s_{1})+\mathcal{D}(s_{2}),\\
 && \mathcal{D}(f s)=f \mathcal{D}s+df \otimes\mathcal{D}s,
 \ee
for any $s, s_{1}, s_{2}\in \Gamma(V), f\in\mathcal{A}=C^{\infty}(N). $
 \ede

 We will consider all the geometric quantities in local sense. Let
 $
 \{s_{\alpha}, 1\leq \alpha \leq m \}
 $
 be the basis of the linear space
 $
 \Gamma(V),
 $
 then
 $
 \{dn^{\mu}\otimes s_{\alpha}, dx^{i}\otimes s_{\beta}, 1\leq\mu\leq p, 1\leq i \leq q, 1\leq\alpha, \beta\leq m \}
 $
 would be the basis of the section space $\Gamma(T^{\ast}N)\otimes \Gamma(V). $
 Therefore, the  left $\mathcal{A}$-module covariant difference-differential derivative may be written locally as
 \be
 \mathcal{D}(s_{\alpha})=-(B_{D, \mu})^{\beta}_{\alpha}dn^{\mu}\otimes s_{\beta}-(B_{C, i})^{\gamma}_{\alpha}dx^{i}\otimes s_{\gamma},
 \ee
 where we have used the Einstein convention of summation of repeated up-down indices and the coefficient $ B_{D, \mu}, B_{C, i}$ are matrix valued.
Just as a generation case of \cite{Wu1}, $B=B_{D, \mu}dn^{\mu}+B_{C, i}dx^{i} $ is the local expression of semi-discrete connection 1-form,
which is matrix valued. Especially, the coefficient $B_{D, \mu}$ is defined on the link $(\vec{n}, \vec{x})\leftrightarrow (\vec{n}+\hat{\mu}, \vec{x})$ and can be written as
\be
B_{D, \mu}((\vec{n}, \vec{x}), (\vec{n}+\hat{\mu}, \vec{x}) ).
\ee
or for simplicity,
\be
B_{D, \mu}(\vec{n}, \vec{n}+\hat{\mu}, \vec{x}) .
\ee

For any section
$
s=f^{\alpha}s_{\alpha},
$
one have
\\
\parbox{12cm}{
\begin{eqnarray*}
\mathcal{D}s &=&df^{\alpha}\otimes s_{\alpha}+f^{\alpha}\mathcal{D}s_{\alpha}\\
                     &= & (d_{D}+d_{C})f^{\alpha}\otimes s_{\alpha}+f^{\alpha}[-(B_{D, \mu})^{\beta}_{\alpha}dn^{\mu}\otimes s_{\beta}-(B_{C, i})^{\gamma}_{\alpha}dx^{i}\otimes s_{\gamma}],\\
                      &= & \Delta _\mu f^{\alpha} dn^{\mu}\otimes s_{\alpha}+\partial_{i}f^{\alpha}dx^{i}\otimes s_{\alpha}\\
                      & & -f^{\alpha}(B_{D, \mu})^{\beta}_{\alpha}dn^{\mu}\otimes s_{\beta}-f^{\alpha}(B_{C, i})^{\gamma}_{\alpha}dx^{i}\otimes s_{\gamma}\\
                      &=&(\Delta _\mu f^{\beta}-f^{\alpha}(B_{D, \mu})^{\beta}_{\alpha} )dn^{\mu}\otimes s_{\beta}+(\partial_{i}f^{\beta}-f^{\alpha}(B_{C, i})^{\beta}_{\alpha}   )dx^{i}\otimes s_{\beta}\\
                      &=&\mathcal{D} _{D_{\mu}}f^{\beta}dn^{\mu}\otimes s_{\beta}+\mathcal{D} _{C_{i}}f^{\beta}dx^{i}\otimes s_{\beta}.
\end{eqnarray*}}\hfill
\parbox{1cm}{\begin{eqnarray}\end{eqnarray}}\\
where
\be
&&\mathcal{D} _{D_{\mu}}f^{\beta}=\Delta _\mu f^{\beta}-f^{\alpha}(B_{D, \mu})^{\beta}_{\alpha},\\
&&\mathcal{D} _{C_{i}}f^{\beta}=\partial_{i}f^{\beta}-f^{\alpha}(B_{C, i})^{\beta}_{\alpha}
\ee
are the discrete and continuous covariant derivative of the vector $f^{\alpha} $ respectively. Similarly, one can also obtain the semi-discrete exterior covariant derivative of the vector $f^{\alpha} $ as
\\
\parbox{12cm}{
\begin{eqnarray*}
\mathcal{D}f^{\beta}&=&\mathcal{D} _{D_{\mu}}f^{\beta}dn^{\mu}+\mathcal{D} _{C_{i}}f^{\beta}dx^{i}\\
                                 &=&(\Delta _\mu f^{\beta}-f^{\alpha}(B_{D, \mu})^{\beta}_{\alpha})dn^{\mu}+(\partial_{i}f^{\beta}-f^{\alpha}(B_{C, i})^{\beta}_{\alpha})dx^{i}.
\end{eqnarray*}}\hfill
\parbox{1cm}{\begin{eqnarray}\end{eqnarray}}

We can also consider the bimodule connection on the $\mathcal{A}$-bimodule $\Gamma(V)$  in this case.
\bpr
If
$\sigma: \Gamma(V)\otimes\Gamma(T^{\ast}N) \rightarrow \Gamma(T^{\ast}N) \otimes\Gamma(V)$ is a $\mathcal{A}$ bimodule map with the property
$$
\sigma (s_{\alpha}\otimes df^{\alpha})=d f^{\alpha}\otimes s_{\alpha}+\Delta _\mu f^{\beta}(B_{D, \mu})_{\beta}^{\alpha}dn^{\mu}\otimes s_{\alpha},
$$
and $\mathcal{D}: \Gamma(V)\rightarrow \Gamma(T^{\ast}N)\otimes \Gamma(V)$ is a left $\mathcal{A}$-module connection, then $ (\Gamma(V), \mathcal{D}, \sigma )$ is a bimodule connection on an $\mathcal{A}$-bimodule $\Gamma(V).$
\epr
\bpro
 Let
 $
 \{s_{\alpha}, 1\leq \alpha \leq m \}
 $
 be the basis of the linear space
 $
 \Gamma(V),
 $
 $f^{\alpha}\in \mathcal{A},$ then one have
 \be
 \sigma(s_{\alpha}\otimes d f^{\alpha})&=& d f^{\alpha}\otimes s_{\alpha}+\Delta _\mu f^{\beta}(B_{D, \mu})^{\alpha}_{\beta}dn^{\mu}\otimes s_{\alpha}\nonumber\\
 &=&[\Delta _\mu f^{\beta}dn^{\mu}+\partial_{i}f^{\beta}dx^{i}]\otimes s_{\beta}+[f^{\alpha}(\vec{n}+\hat{\mu}, \vec{x})-\nonumber\\
 & & f^{\alpha}(\vec{n}, \vec{x})](B_{D, \mu})^{\beta}_{\alpha}dn^{\mu}\otimes s_{\beta},
 \ee
 \be
\mathcal{D}s_{\alpha}\cdot f^{\alpha}&=& [-(B_{D, \mu})^{\beta}_{\alpha}dn^{\mu}\otimes s_{\beta}-(B_{C, i})^{\gamma}_{\alpha}dx^{i}\otimes s_{\gamma}]\cdot f^{\alpha}\nonumber\\
&=& -f^{\alpha}(\vec{n}+\hat{\mu}, \vec{x})(B_{D, \mu})^{\beta}_{\alpha}dn^{\mu}\otimes s_{\beta}-f^{\alpha}(B_{C, i})^{\gamma}_{\alpha}dx^{i}\otimes s_{\gamma}.
\ee
Therefore, one have
\be
&&\mathcal{D}s_{\alpha}\cdot f^{\alpha}+\sigma(s_{\alpha}\otimes d f^{\alpha})=f^{\alpha}\mathcal{D}s_{\alpha}+df^{\alpha}\otimes s_{\alpha},\\
&&\mathcal{D}(s_{\alpha}\cdot f^{\alpha})=\mathcal{D}s_{\alpha}\cdot f^{\alpha}+\sigma(s_{\alpha}\otimes d f^{\alpha}).
\ee
Hence, $ (\Gamma(V), \mathcal{D}, \sigma )$ is a bimodule connection on an $\mathcal{A}$-bimodule $\Gamma(V).$
\qed
\epro

It should be noted that although we have obtained the bimodule connection on an $\mathcal{A}$-bimodule $\Gamma(V)$, we will just use the left  $\mathcal{A}$-module  connection $\mathcal{D}$ all over this paper.
\subsection{Noncommutative semi-discrete curvature }
In this subsection, we will obtain the noncommutative semi-discrete curvature according to Def. 2.2.2.

From the Def. 4.2.1, one know that a noncommutative difference-differential connection $\mathcal{D}$
 $$
 \mathcal{D}: \Gamma(V)\rightarrow \Gamma(T^{\ast}N)\otimes \Gamma(V)
 $$
is of the form
$$
\mathcal{D}s =(\Delta _\mu f^{\beta}-f^{\alpha}(B_{D, \mu})^{\beta}_{\alpha} )dn^{\mu}\otimes s_{\beta}+(\partial_{i}f^{\beta}-f^{\alpha}(B_{C, i})^{\beta}_{\alpha}   )dx^{i}\otimes s_{\beta}.
$$
By Def. 2.2.2., one have the curvature 2-form as,
$$
\mathcal{D}^{2}:  \Gamma(V)\rightarrow \Omega^{2}(N)\otimes\Gamma(V).
$$
\bth
If $s=f^{\alpha}s_{\alpha}\in\Gamma(V)$ is a section of the semi-discrete vector bundle $V(N, GL(m, \mathds{R}))$, the curvature $\mathcal{D}^{2}$ is of the form:
$$
\mathcal{D}^{2}s=-\frac{1}{2}[f^{\alpha}(F_{\mu, \nu})^{t}_{\alpha}dn^{\mu}\wedge dn^{\nu}+f^{\alpha}(F_{i, j})^{t}_{\alpha}dx^{i}\wedge dx^{j}+f^{\alpha}(F_{\mu, i})^{t}_{\alpha}dn^{\mu}\wedge dx^{i}]\otimes s_{t},
$$
where
\be
F_{\mu, \nu}&=&\Delta _\mu B_{D,\nu}-\Delta _\nu B_{D,\mu}+B_{D,\mu} B_{D,\nu}(\vec{n}+\hat{\mu},\vec{ x})\nonumber\\
& & -B_{D,\nu}B_{D,\mu}(\vec{n}+\hat{\nu},\vec{ x}),\\
F_{i, j}&=&\partial_{i}B_{C,j}-\partial_{j}B_{C,i}+B_{C,i}B_{C,j}-B_{C,j}B_{C,i},\\
F_{\mu, i}&=&\Delta _\mu B_{C,i}-\partial_{i} B_{D,\mu}+B_{D,\mu}B_{C,i}(\vec{n}+\hat{\mu},\vec{ x})-B_{C,i}B_{D,\mu}
\ee
are the noncommutative difference curvature form, the ordinary continuous curvature form and the difference-differential curvature form respectively.
\eth
\bpro
If $s=f^{\alpha}s_{\alpha}$, then
$
\mathcal{D}s =(\Delta _\mu f^{\beta}-f^{\alpha}(B_{D, \mu})^{\beta}_{\alpha} )dn^{\mu}\otimes s_{\beta}+(\partial_{i}f^{\beta}-f^{\alpha}(B_{C, i})^{\beta}_{\alpha}   )dx^{i}\otimes s_{\beta}.
$
By means of the property of $\mathcal{D}$:
$$
\mathcal{D}(\omega \otimes_{\mathcal{A}}m)=(-1)^{n}\omega \mathcal{D}(m)+d(\omega)\otimes_{\mathcal{A}}m,
$$
one can obtain the result by directly apply $\mathcal{D}$ to $ \mathcal{D}s.$\qed
\epro
\numberwithin{equation}{section}
\section{Applications in continuous, discrete and semi-discrete integrable systems}
\label{sec:5}
\hspace{0.5 cm}
 In this section, we will use the theory of noncommutative connection and curvature to  discuss some
differential, difference and difference-differential equations. One can see that if the curvature is zero, one can get the corresponding integrable equations.
 A nonlinear equation admits a zero curvature representation if it is equivalent to the compatibility condition of a pair of auxiliary linear problems. For the Partial differential(DD), differential-difference $(D\Delta)$and
difference-difference $(\Delta\Delta)$ equations, the corresponding auxiliary problems are listed as:
  $$
 \begin{array}{ll}
 DD:                & \partial_{x_{i}}\vec{\Psi}=\vec{\Psi}B_{C,i}(\vec{x}),~~~~ \partial_{x_{j}}\vec{\Psi}=\vec{\Psi} B_{C,j}(\vec{x})~~~~ \Rightarrow\\
                      & \partial_{x_{i}}B_{C,j}(\vec{x})-\partial_{x_{j}}B_{C,i}(\vec{x})+B_{C,i}(\vec{x})B_{C,j}(\vec{x})-
 B_{C,j}(\vec{x})B_{C,i}(\vec{x})=0,\\
 D\Delta:         &\partial_{x}\vec{\Psi_{n}}=\vec{\Psi_{n}}U(\vec{n}, \vec{x}), ~~~~ \Delta_{\mu}\vec{\Psi_{n}}=\vec{\Psi_{n}}V(\vec{n}, \vec{x})~~~~ \Rightarrow \\
                    &\Delta _{\mu}U(\vec{n}, \vec{x})-\partial_{x}V(\vec{n}, \vec{x})+V(\vec{n}, \vec{x})U(\vec{n}+\hat{\mu}, \vec{x})-U(\vec{n}, \vec{x})V(\vec{n}, \vec{x})=0,\\
 \Delta\Delta:  & \Delta_{\nu}\vec{\Psi_{n}}= \vec{\Psi_{n}}B_{D, \nu}(\vec{n}),~~~~\Delta_{\mu}\vec{\Psi_{n}}= \vec{\Psi_{n}}B_{D, \mu}(\vec{n})~~~~ \Rightarrow\\ &\Delta_{\mu}B_{D, \nu}(\vec{n})-\Delta_{\nu}B_{D, \mu}(\vec{n})+B_{D, \mu}(\vec{n})B_{D, \nu}(\vec{n}+\hat{\mu})-B_{D, \nu}(\vec{n})B_{D, \mu}(\vec{n}+\hat{\nu})=0.
 \end{array}
$$
 They are just the three types of curvatures we have derived in Theom. 4.3.1.

 We will give out three concrete realization of the  expression of semi-discrete connection 1-form $B=B_{D, \mu}dn^{\mu}+B_{C, i}dx^{i} $. One can see that the zero-curvature conditions are equivalent to three type of integrable equations.

(1) Let $B_{D, \mu}=0, B_{C, i}\in GL(2, \mathds{R}), i=1, 2. $  i.e., the semi-discrete connection 1-form
 $B=B_{D, \mu}dn^{\mu}+B_{C, i}dx^{i}=B_{C, 1}dx+ B_{C, 2}dt, $ with the coefficients as:
 $$
 B_{C, 1}(x, t)=\left(
\begin{array}{cc}
0  & u^{\ast}(x, t) \\
-u(x, t)  & 0
\end{array}
\right),
B_{C, 2}(x, t)=\left(
\begin{array}{cc}
-i \left| u(x, t)\right|^{2}      & i\frac{du(x, t)}{dx}  \\
i\frac{du^{\ast}(x, t)}{dx}  & i \left| u(x, t)\right|^{2}
\end{array}
\right),
$$
The zero-curvature condition  is
\begin{eqnarray}
F_{1 2} &:=& \partial_{t}B_{C, 1}(x, t)-\partial_{x} B_{C, 2}(x, t)+B_{C, 2}(x, t)B_{C, 1}(x, t)-B_{C, 1}(x, t) B_{C, 2}(x, t)\nonumber\\
  &=&
 \left(
\begin{array}{cc}
 Z_{11}&  0\\
0   & Z_{22}
\end{array}
\right)\nonumber\\
&=& 0.
\end{eqnarray}
where
$$
Z_{11}=\frac{\partial u}{\partial t}+i \frac{\partial^{2} u}{\partial x^{2}}+2i \left| u\right|^{2}u,~~~
Z_{22}=\frac{\partial u^{\ast}}{\partial t}-i \frac{\partial^{2} u^{\ast}}{\partial x^{2}}-2i \left| u\right|^{2}u^{\ast}.
$$
It can be verified that equation
$$
F_{1 2}=0
$$
is equivalent to the following famous nonlinear $Schr\ddot{o}dinger$ equation,
\begin{equation}
\frac{\partial u(x, t)}{\partial t}+i \frac{\partial^{2} u(x, t)}{\partial x^{2}}+2i \left| u(x, t)\right|^{2}u(x, t)=0.
\end{equation}

(2) Let the $ B_{C, i}, B_{D, \mu}\in GL(2, \mathds{R}), i=\mu=1. $  i.e., the semi-discrete connection 1-form
 $B=B_{D, \mu}dn^{\mu}+B_{C, i}dx^{i}=B_{C, 1}dx+ B_{D, 1}dn, $ with the coefficients as
 $$
 B_{C, 1}(x, n)=\left(
\begin{array}{cc}
-1  & \frac{\gamma}{ik}e^{i\theta_{n}} \\
  \frac{\gamma}{ik}e^{-i\theta_{n}} & -1
\end{array}
\right),
B_{D, 1}(x, n)=\left(
\begin{array}{cc}
-1+e^{-\frac{i}{2}(\theta_{n+1}-\theta_{n})}      & ik  \\
ik    &  -1+e^{\frac{i}{2}(\theta_{n+1}-\theta_{n})}
\end{array}
\right),
$$
The zero-curvature condition  is
\begin{eqnarray}
F_{1 2} &:=& \bigtriangleup_{1}B_{C, 1}(x, n)-\partial_{x} B_{D, 1}(x, n)+B_{D, 1}(x, n)B_{C, 1}(x, n+1)-B_{C, 1}(x, n) B_{D, 1}(x, n)\nonumber\\
  &=&
 \left(
\begin{array}{cc}
 Z_{11}&  0\\
0   & Z_{22}
\end{array}
\right)\nonumber\\
&=& 0.
\end{eqnarray}
where
\begin{eqnarray*}
&&Z_{11}=-\frac{i}{2}e^{-\frac{i}{2}(\theta_{n+1}-\theta_{n})}\partial_{t}(\theta_{n+1}-\theta_{n})+
e^{-\frac{i}{2}(\theta_{n+1}-\theta_{n})}
-\gamma e^{-i(\theta_{n+1})},\\
&&Z_{22}=\frac{i}{2}e^{\frac{i}{2}(\theta_{n+1}-\theta_{n})}\partial_{t}(\theta_{n+1}-\theta_{n})+\gamma e^{-i\theta_{n}}
-\gamma e^{i(\theta_{n+1})}.
\end{eqnarray*}
It can be verified that equation
$$
F_{1 2}=0
$$
is equivalent to the following difference-differential  Equation,
\begin{equation}
\partial_{t}\theta_{n+1}-\partial_{t}\theta_{n}=
\gamma\sin\frac{1}{2}(\theta_{n+1}+\theta_{n}),
\end{equation}
where $ n $ is an integer variable specifying
location in the lattice, $ t $ is the continuous temporal variable.
which is an integrable discretization model of the Sine-Gordon equation.

 (3) Let the $ B_{C, i}=0,$ $B_{D, \mu}\in GL(2, \mathds{R}), \mu=1,2.$ i.e., the semi-discrete connection 1-form
 $B=B_{D, \mu}dn^{\mu}+B_{C, i}dx^{i}=B_{D, 1}dm+ B_{D, 2}dn, $ with the coefficients as
 $$
 B_{D, 1}(m, n)=\left(
\begin{array}{cc}
\lambda-1  & \frac{\lambda}{u_{m, n-1}} \\
  u_{m, n} & -2
\end{array}
\right),
B_{D, 2}(m, n)=\left(
\begin{array}{cc}
\lambda-1+\frac{u_{m, n}}{u_{m-1, n}}      & \frac{\lambda}{u_{m-1, n}}  \\
u_{m, n}     &  -1
\end{array}
\right),
$$
The zero-curvature condition  is
\begin{eqnarray}
F_{1 2} &:=& \bigtriangleup_{2}B_{D, 1}(m, n)-\bigtriangleup_{1} B_{D, 2}(m, n)+B_{D, 2}(m, n)B_{D, 1}(m, n+1)-B_{D, 1}(m, n) B_{D, 2}(m+1, n)\nonumber\\
  &=&
 \left(
\begin{array}{cc}
\lambda[\frac{u_{m, n}}{u_{m-1, n}}+\frac{u_{m, n+1}}{u_{m-1,
n}}-\frac{u_{m+1, n}}{u_{m, n}}-
\frac{u_{m+1, n}}{u_{m, n-1}}] &  0\\
0   & 0
\end{array}
\right)\nonumber\\
&=& 0.
\end{eqnarray}
It can be verified that equation
$$
F_{1 2}=0
$$
is equivalent to the following partial difference  Equation,
\begin{equation}
q_{m+1, n}-2q_{m, n}+q_{m-1, n}=\ln\frac{e^{q_{m,n+1}-q_{m, n}}+1}{e^{q_{m, n}-q_{m, n-1}}+1},
\end{equation}
with
$$
u_{m, n}=e^{q_{m, n}}.
$$
which is an integrable discretization model of the famous Toda lattice equation\cite{s31}.
\section{Conclusions}
\label{sec:7} \hspace{0.5 cm} In this paper, we have proposed  a more general realization of the noncommutative connection and curvature on the discrete bundle from the viewpoint of algebra. The whole procedure is based on a concrete deformed derivation-based differential calculus. As is clear, the algebra we have considered is actually commutative. However, the module that has been used is not a bimodule. We have just consider the left $\mathcal{A}$-module $\Omega^{\ast}(\mathcal{A}) $.  Equivalently, there also exist the corresponded theory of right $\mathcal{A}$-module $\Omega^{\ast}(\mathcal{A}) $. It should be noted that the theory is not just a parallel of the ordinary commutative differential geometry.

We have just considered the application of the theory to integrable systems. The theory of discrete principle bundle, the associated vector bundle, the connection and curvature might also be applied to other  mathematical physics areas such as lattice gauge theory.

We have not consider the metric aspects of the noncommutative geometry at all. Therefore, it is also possible to use the theory to discuss spectral triple on the noncommutative side.

Recently, many noncommutative soliton equations have appeared in discussing noncommutative field theories.
In this case, the underlying noncommutative algebra $\mathcal{A} $ is the Moyal  algebra. The integrability and noncommutative zero-curvature representation of these equations are deserved further studied in the future.

{\bf Acknowledgment:}
The project was supported by National Natural
Science Foundation of China (Grant No. 10801045).
%
%
%

\end{document}